\documentclass[12pt,preprint]{aastex}

\newcommand{\lea}{{\>\rlap{\raise2pt\hbox{$<$}}\lower3pt\hbox{$\sim$} \>}}
\newcommand{\gea}{{\>\rlap{\raise2pt\hbox{$>$}}\lower3pt\hbox{$\sim$} \>}}

\begin{document}

\title{A Comparison of Methods for Determining the Age Distribution of Star Clusters: Application to the Large Magellanic Cloud}

\author{Rupali Chandar,$^{1}$ Bradley C.\ Whitmore,$^{2}$ and S.\ Michael Fall$^{2}$}
\affil{$^1$ Department of Physics and Astronomy, University of Toledo, Toledo, OH 43606\\
$^2$ Space Telescope Science Institute, 3700 San Martin Drive,
Baltimore, MD 21218}

\begin{abstract}
The age distribution of star clusters in nearby galaxies plays a crucial role in evaluating the lifetimes and disruption mechanisms of the clusters. Two very different results have been found recently for the age distribution  $\chi(\tau)$ of clusters in the Large Magellanic Cloud (LMC).  We found that $\chi(\tau)$ can be described approximately by a power law $\chi(\tau) \propto \tau^{\gamma}$, with $\gamma\approx-0.8$, by counting clusters in the mass-age plane, i.e., by constructing $\chi(\tau)$ directly from mass-limited
samples. Gieles \& Bastian inferred a value of $\gamma\approx0$, based on the slope of the relation between the maximum mass of clusters in equal intervals of $\log\tau$, hereafter the $M_{\max}$ method, an indirect
technique that requires additional assumptions about the upper end of the mass function. However, our own analysis shows that the $M_{\max}$
method gives a result consistent with our direct counting method for clusters in the LMC, namely $\chi(\tau)\propto\tau^{-0.8}$ for $\tau \lea 10^9$~yr.
The reason for the apparent discrepancy is that our analysis includes many intermediate and high-mass clusters ($M>1.5\times10^{3}~M_{\odot}$), 
which formed recently ($\tau \lea 10^7$~yr), and which are known to exist 
in the LMC, whereas Gieles \& Bastian are missing such clusters. We compile recent results from the literature showing that the age distribution of young star clusters in more than a dozen galaxies, including dwarf and giant
galaxies, isolated and interacting galaxies, irregular and spiral galaxies, 
has a similar declining shape. We interpret this approximately ``universal'' shape as due primarily to the progressive disruption of star clusters over their
first $\sim\mbox{few}\times10^8$~yr, starting soon after formation, and 
discuss some observational and physical implications of this early disruption for stellar populations in galaxies.
\end{abstract}


\section{INTRODUCTION}
Many and possibly all stars form in clusters, which are then dispersed into the general field population by a variety of physical processes. The imprint of these processes is reflected in the mass and age distributions of a population of star clusters, $\psi(M)\equiv dN/dM$ and $\chi(\tau)\equiv dN/d\tau$, and more generally in their bivariate mass-age distribution $g(M,\tau)$.
Information on the physical processes that affect clusters can be gleaned by comparing the $g(M,\tau)$ distributions in different galaxies. In this paper and related works, we take a ``cluster'' to be any concentrated aggregate of stars, with a density much higher than that of the surrounding stellar field,
whether or not it is gravitationally bound, since the latter is nearly impossible to determine from observations, particularly for clusters younger than about ten internal crossing times.

We previously determined the $g(M,\tau)$, $\psi(M)$, and $\chi(\tau)$ distributions of star clusters in the merging Antennae galaxies and in the more typical Magellanic Clouds. In all three galaxies, we found that $\psi(M)$ and $\chi(\tau)$ can be approximated by power laws and are roughly independent of one another for young clusters with $\tau \lea 10^8$--$10^9$~yr; thus $g(M,\tau) \propto \psi(M) \chi(\tau) \propto M^{\beta}\tau^{\gamma}$, with $\beta \approx-2$ and $\gamma\approx-1$ (Zhang \& Fall 1999; Fall et~al.\ 2005, hereafter FCW05; Whitmore et~al.\ 2007, hereafter WCF07; Fall et~al.\ 2009, hereafter FCW09; Chandar et~al.\ 2010, hereafter CFW10). We obtained these results by counting clusters in relatively narrow bands of age and mass to determine the $\psi(M)$ and $\chi(\tau)$ 
distributions, and refer to this as the direct counting method.

Gieles \& Bastian (2008; hereafter GB08) suggested that the age distribution of star clusters in the Magellanic Clouds has a flat rather than a declining shape. In our notation, their result can be expressed as $g(M,\tau) \propto M^{\beta} \tau^{\gamma}$, with $\beta\approx-2$ and $\gamma\approx0$.
GB08 plotted the mass of the most massive cluster as a function of age, in the form log~$M_{\max}$ vs.\ $\log\tau$, referred to here as the $M_{\max}$ method, to {\em infer} the shape of the age distribution.
Unlike the direct counting method, which uses all of the available data, the $M_{\max}$ method relies on only a handful of clusters that reside at the upper envelope of the two-dimensional $M$--$\tau$ distribution, and is sensitive to fluctuations due to small-number statistics and the accidental presence or absence of only a few clusters. In addition, because this technique does not measure $\chi(\tau)$ directly, it depends on several assumptions: that the value of $\beta$ does not change over time, that there is no upper mass cutoff $M_C$ which would alter the expected distribution
for the most massive clusters, and that obscuration does not significantly impact the luminosities of massive young clusters.

We focus here on clusters in the Large Magellanic Cloud (LMC).\footnote{The SMC has formed fewer clusters than the LMC, and the low numbers in
current catalogs result in poor statistics which can have a significant impact on $M_{\max}(\tau)$. In addition, current samples of clusters in the SMC are known to be incomplete, particularly at ages $\tau \lea 10^7$~yr
(see the Appendix in CFW10), which can bias the age distribution.} The two different values of $\gamma$ mentioned above were determined from two entirely {\em different} methodologies, applied to the {\em same} catalog of star clusters. The plan for the remainder of this paper is the following:
In Section~2, we present Monte Carlo simulations of the two models for $g(M,\tau)$ described above. In Section~3, we present the age distribution of
clusters in the LMC from the direct counting method, and find $\gamma \approx-0.8$. In Section~4, we revisit the $M_{\max}$ method using our own mass and age estimates of clusters in the LMC, and show that this method also gives results consistent with $\gamma\approx-0.8$. In Section~5, we check whether there is  an upper mass limit or cutoff $M_C$
for clusters in the LMC, and discuss the impact on the $M_{\max}$ method if such a cutoff exists. In Section~6, we compile evidence that the age distribution of star clusters in more than a dozen different galaxies has a 
nearly``universal'' shape, and in Section~7 we summarize our main conclusions.

\section{MONTE CARLO SIMULATIONS AND AGE DISTRIBUTIONS FOR TWO DIFFERENT MODELS}

The fundamental difference between our result and that of GB08 for clusters in the LMC is the value of $\gamma$, the power-law index of $\chi(\tau)$.
To illustrate this difference, we simulate $g(M,\tau)$ for two different models:
a declining $\chi(\tau)$ distribution (Model~A) and a flat  $\chi(\tau)$ distribution (Model~B). A Monte Carlo realization of the $M$--$\tau$ plane predicted by each model is shown in Figure~1, with $g(M,\tau) \propto M^{-1.8}\tau^{-0.8}$ for Model~A (top panel) and $g(M,\tau) \propto M^{-2.0}\tau^{0.0}$ for Model~B (bottom panel). The specific values of $\beta$ and $\gamma$ are motivated by the results presented in CFW10 for Model~A,
and from the results in GB08 for Model~B. The number of simulated clusters above the solid line (which represents $M_V=-4.0$) for both models approximately matches the observed number of clusters brighter than this limit in the Hunter et~al.\ sample, described in Section~3.

Several important differences can be seen in the predicted $M$--$\tau$ diagrams shown in Figure~1, which contain all of the statistical information on the masses and ages of a population of clusters. First,  the upper envelope of data points increases gradually with age for Model~A, but rapidly for Model~B; the slope of this upper envelope is used in the $M_{\max}$ method to infer $\gamma$, as described in Section~4.1.
Second, Model~B predicts very few clusters younger than $\tau \lea10^7$~yr, and that any such clusters will have relatively low masses, with none reaching $M\sim10^4~M_{\odot}$ and only a few reaching $\sim\!10^3~M_{\odot}$. Model~A, on the other hand, predicts the formation of many clusters with $M\sim10^3~M_{\odot}$, at least a few with $M\sim10^4~M_{\odot}$, and possibly one with $M\sim10^5~M_{\odot}$, in the last $\tau \lea 10^7$~yr. Third, for Model~A the number of clusters increases gradually with age (in equal bins of $\log\tau$) for a  fixed interval in mass, while for Model~B the age distribution is uniform in $\tau$, and hence clusters ``pile up'' at older ages in $\log\tau$.

Figure~2 shows age distributions constructed directly from the $g(M,\tau)$ distributions of Models~A and B, by counting clusters in the indicated intervals of $\log\tau$ and $\log~M$, i.e., for mass-limited samples (the age distributions inferred from the $M_{\max}$ method for these simulations are discussed in Section~4.1). We use the same bins as for the LMC data in Section~3, although some data points are missing where Model~B predicts no clusters for the youngest age bins. Figure~2 confirms that, by construction, $\gamma\approx-0.8$ for Model~A and $\gamma\approx0.0$ for Model~B. The strong differences predicted by Models~A and B for the age distribution, which are obvious even in the $M$--$\tau$ diagram, should make it easy to determine which model gives a better description of young clusters in the LMC.

The bivariate $g(M,\tau)$ distribution, and the $\chi(\tau)$ and $M_{\max}$ relations that follow from it, are shaped by the difference between the formation and disruption rates of the clusters. In the LMC, the age distribution is almost a pure reflection of the disruption rate, because the star formation rate has been nearly constant (i.e., varied by less than a factor of two) over the last several Gyr (Harris \& Zaritsky 2009). The two different values of $\gamma$ suggested above therefore, have very different
physical implications for the disruption of star clusters in the LMC. Model~A, with $\gamma\approx-0.8$, leads to a picture where star clusters in the LMC 
are relatively fragile, with most falling apart within a few hundred Myr of 
birth, regardless of their initial mass (CFW10). Model~B, with $\gamma \approx0$, leads to a scenario where clusters in the LMC are incredibly
durable, and once formed are difficult to destroy. Further observational and physical implications for star formation and cluster evolution are discussed in Section~6.

\section{THE AGE DISTRIBUTION OF STAR CLUSTERS IN THE LMC FROM THE \boldmath{$M$}--\boldmath{$\tau$} DIAGRAM}

Here we summarize our results from CFW10 for the age distribution of star clusters in the LMC using our direct counting method. We estimated the age of each of the 854 LMC clusters in the Hunter et~al.\ (2003) sample by performing a least $\chi^2$ fit comparing their \textit{UBVR} magnitudes
with predictions from the Bruzual \& Charlot (2003) models for simple stellar populations, assuming a metallicity $Z=0.008$ (40\% of the solar value), a Salpeter initial mass function (IMF), and a Galactic-type extinction curve (Fitzpatrick 1999). We estimated the mass of each cluster from the $V$~band luminosity (corrected for extinction) and the age-dependent mass-to-light ratios ($M/L_V$) predicted by the Bruzual \& Charlot models, assuming a distance modulus to the LMC of 18.5 (Alves 2004). We found uncertainties of $\approx0.3$--0.4 in both $\log\tau$ and $\log~M$ (CFW10).
More details about the data, dating procedure, and uncertainties in the age and mass estimates are given in CFW10. The resulting $M$--$\tau$ diagram of clusters in the LMC is shown in the top panel of Figure~3.

This $M$--$\tau$ diagram shows a number of small-scale features, including gaps and ridges at specific ages, which result from well-known artifacts that arise during the dating procedure (see discussion in CFW10). These features do not impact the broad distribution of points in this plane, which is of interest here. The basic results for the age distribution of clusters in the LMC are immediately obvious from the $M$--$\tau$ diagram, and can be compared with the predictions from Models~A and B. First, the upper envelope of data points in the $M$--$\tau$ plane increases gradually, not rapidly, with age. This will be discussed in more detail in Section~4.2. Second, we find several clusters in the LMC with masses $M \gea 10^4~M_{\odot}$ and ages $\tau \lea 10^7$~yr. Third, the number of clusters in equal bins of $\log\tau$ above a given mass increases slowly, not rapidly, with age. All of these features are quite similar to predictions from Model~A, but very different from those from Model~B.

More quantitatively, the $\chi(\tau)$ distributions resulting from our direct counting method and presented in CFW10 for three different intervals of mass are reproduced in the bottom panel of Figure~3. We included as many clusters in the $M$--$\tau$ plane as possible, but stopped counting before we reached $M_V=-4.0$ (shown as the solid line in the top panel of Figure~3), where the data become significantly incomplete. All three mass-limited distributions have a declining shape. In CFW10 we found similar results
for $\chi(\tau)$ if we used the cluster age estimates from Hunter et~al.\ (2003) instead of our own. The age distributions can be approximated by a power law, $\chi(\tau) \propto \tau^{\gamma}$, with $\gamma=-0.8\pm0.2$,
very similar to the predictions from Model~A. Figure~10 in Parmentier \& de Grijs (2008) shows a similar declining shape for $\chi(\tau)$ for mass ranges that are similar to those used here, based  on their own age estimates and independent analysis of the Hunter et~al.\ (2003) sample,
although they give a different physical interpretation for this shape. We conclude, therefore, that Model~A, with $g(M,\tau) \propto M^{-1.8} \tau^{-0.8}$, provides a good approximation to the properties of clusters in the LMC over the plotted range of masses and ages (i.e., $\tau \lea 10^7(M/10^2 M_{\odot})^{1.3}$~yr).

\section{THE AGE DISTRIBUTION FROM THE \boldmath{$M_{\max}$} METHOD}

\subsection{Predictions}

The $M_{\max}$ method provides an alternative estimate of $\chi(\tau)$, based on a small subsample of clusters. We start by deriving the scaling relation for the expected maximum mass $M_{\max}$ as a function of age $\tau$ in equal logarithmic bins $\Delta \log\tau$ for a power-law mass function. The condition for finding {\em no} clusters with masses above $M_{\max}$ is
\begin{equation}
\int_{M_{\max}(\tau)}^{\infty} \frac{\partial^2N}{\partial M \partial\log\tau} dM \Delta\log\tau~~ =~~ \mbox{const}~~ \sim~~ 1.
\end{equation}
We rewrite the integrand of this equation using the mass-age distribution defined previously: 
$\partial^2N/\partial M \partial\log\tau\propto\tau g(M,\tau) = \tau \psi(M) \chi(\tau)$, where the last expression is based on the assumption that the
mass and age distributions are independent of one another. For constant $\Delta\log\tau$, Equation~(1) then becomes
\begin{equation}
\tau \chi(\tau) \int_{M_{\max}(\tau)}^{\infty} \psi(M)dM~~ =~~ \mbox{const}^{\prime}.
\end{equation}
At this point, we approximate the mass and age distributions by power laws:
$\psi(M) \propto M^{\beta}$ and $\chi(\tau) \propto \tau^{\gamma}$.  
(This is Model~3 from FCW09 and CFW10.) We then have
\begin{equation}
M_{\max} \propto \tau^{\delta},~~~~~ \mbox{with}~~~\delta=-(1+\gamma)/(1+\beta)~~~~~~~  \mbox{for $\beta$ $< -1$.}
\end{equation}
Thus, $\delta$ is the slope of the $\log~M_{\\max}$ vs.\ $\log\tau$ relation. This expression for $M_{\max}$ is equivalent to Equation~(10) in GB08. For the two models of interest here, we then have
%\begin{equation}
%\mbox{Model~A}:~~~ \delta={0.25}~~~~~ \mbox{for}~ \beta=-1.8~ \mbox{and}~ \gamma=-0.8,
%\end{equation}
%\begin{equation}
%\mbox{Model~B}:~~~ \delta={1.0}~~~~~ \mbox{for}~ \beta=-2.0~ \mbox{and}~ \gamma=0.0.
%\end{equation}
\begin{eqnarray}
\mbox{Model~A}:\quad \delta={0.25}\quad &\mbox{for}\ \beta=-1.8 &\mbox{and}\ \gamma=-0.8,\\
\mbox{Model~B}:\quad \delta={1.0}\enspace\quad &\mbox{for}\ \beta=-2.0 &\mbox{and}\ \gamma=0.0.
\end{eqnarray}
These equations only apply for bins that are equal in $\log\tau$; $M_{\max}$ will have a different dependence on $\tau$ if the binning is not logarithmic.

Note that the $M_{\max}$ relationship depends only on the upper envelope of data points in the $M$--$\tau$ plane, and hence on the {\em assumed} (not measured) shape of the upper end of the mass function.  The $M_{\max}$ relations for Models~A and B are shown as the dashed and dotted lines in Figure~1. The normalization along the ordinate (at $\log\tau=6.0$) for Model~A is the mean value from our Monte Carlo simulations, while that for Model~B is taken from GB08 and provides a good match to the simulations.
Typical observational uncertainties of $\approx\!0.2$ in $\beta$ and $\gamma$ give an uncertainty of $\approx\!0.3$ in the predicted value of $\delta$, based on Equation~(3) and propagation of errors.

Our assumption that $\beta$ is constant is based on our empirical study of the $g(M,\tau)$ distribution for the LMC clusters, but it also has some theoretical support. The removal of interstellar material (ISM) by feedback from massive stars on timescales $\tau\lea 10^7$~yr, can unbind many protoclusters (e.g., Hills 1980). Fall et~al.\ (2010) showed that $\beta$ is nearly preserved if the feedback is momentum-driven, and if the protoclusters
initially have approximately constant mean surface density, as indicated by observations of star-forming clumps within molecular clouds. Following this, clusters continue to lose mass due to stellar evolution, which can unbind those clusters that are weakly bound by the prior removal of their ISM. We argued in FCW09 that if the concentration parameters of clusters are uncorrelated with their masses, a large fraction of them could also be disrupted in the period $10^7~\mbox{yr} \lea \tau \lea 10^8$~yr, without changing $\beta$.

\subsection{Comparison with Observations}

Here, we determine the exponent $\delta$ in Equation~(3) for clusters in the LMC from the $M_{\max}$ method. In particular, we find the most massive cluster in bins of $\Delta\log\tau=1$ starting at $\log\tau=6$, using our mass and age estimates of clusters in the Hunter et~al.\ (2003) sample.
The result is shown in Figure~4, and gives $\delta\approx0.3$, based on a simple linear fit. We find a similar value of $\delta$ if the third most massive
cluster is used instead. A slope of $\delta=0.3$ implies $\gamma\approx -0.8$
for $\beta=-1.8$, according to Equation~(3). GB08 plotted the most massive cluster in bins of $\Delta\log\tau=0.5$, using the ages estimated by Hunter et~al.\ (2003), and found $\delta\approx1$ for star clusters in the LMC. From this and the assumption that $\beta=-2.0$, they inferred that the age distribution is flat, with $\gamma\approx0$.

A comparison between the $M_{\max}$ relation found here and that from GB08 is also shown in Figure~4, and reveals that the results are quite similar, 
despite the significantly steeper slope claimed by GB08 ($\delta\approx1$ vs.\ $\delta\approx0.3$). The critical difference comes from a single data point, 
where GB08 find $M_{\max}\approx1.5\times10^3~M_{\odot}$ for clusters with $\log(\tau/\mbox{yr}) < 7$, which is lower by a factor of $\sim\!30$ than found here, and is responsible for nearly all of the weight in their fit.

The GB08 estimate of $\delta$ and hence $\gamma$ can only be correct if the LMC has not formed any massive young clusters, i.e., there should be no clusters in the triangular region above the dotted line in Figure~4, a region where our analysis clearly places clusters. It is well known, however, that the LMC {\em has} formed clusters more massive than $1.5\times10^3~M_{\odot}$  in the last $10^7$~yr. The most famous example is R136 in the 30~Doradus nebula, which has an age of $\tau\approx3\times10^6$~yr and a mass of $M\sim10^5~M_{\odot}$ (e.g., McLaughlin \& van der Marel 2005), and there are several other young clusters with masses $\sim\!10^4~M_{\odot}$ (e.g., H88$-$267, SL360, NGC~2100). Therefore, the GB08 result for $\delta$, and hence for $\gamma$, is incorrect.

Our own dating analysis finds 36 clusters in the LMC that are more massive than $1.5\times10^{3}~M_{\odot}$ and younger than $10^7$~yr. We performed an independent check of our age estimates for these clusters by locating them in two sets of publicly available $H\alpha$ images: (1)~low-resolution images ($0.8\arcmin~\mbox{pix}^{-1}$) from the SHASSA survey\footnote{The continuum-subtracted images are available from  URL 
http://amundsen.swarthmore.edu/\#Specifications.} which cover the entire LMC, and (2)~higher-resolution images ($2.3\arcsec~\mbox{pix}^{-1}$) from the Magellanic Cloud Emission Line Survey (MCELS)\footnote{The
images are available from http://www.ctio.noao.edu/~mcels/.} (the images currently available from MCELS do not cover the entire LMC), and find that $\approx\!65$\% are HII regions. This gives a minimum fraction of our sample
that is younger than $10^7$~yr, since we may be missing faint H$\alpha$ emission from some clusters and some clusters clear their natal gas on timescales shorter than $10^7$~yr. We find that the clusters that are HII regions all have estimated masses higher than $1.5\times10^3~M_{\odot}$,
even if we make no correction for extinction. In fact, three such clusters have estimated masses of $M\gea10^4~M_{\odot}$ (uncorrected for extinction), and several others are just below $10^4~M_{\odot}$, validating our mass-age estimates.

Our result for $\delta$ is corroborated by the independent mass and age estimates for LMC clusters presented by McLaughlin \& van der Marel (2005).
Their sample was selected to include some of the brightest known clusters at different ages, although its completeness has not been rigorously assessed.
The most massive cluster in each $\Delta\log\tau=1$ bin starting at $\log\tau=6.0$ from their study is shown as the large, solid circles in Figure~5.
The slope of this $M_{\max}$ relation is $\delta\approx0.3$, nearly identical
to the value we found for the Hunter et~al.\ sample (with our age estimates).

These results show that the youngest data point in the $M_{\max}$ relation found by GB08, the only one that differs in any appreciable way from those found here, is artificially low. This is due to a systematic bias in the ages estimated by Hunter et~al.\ (2003), such that their technique can assign somewhat older ages to the youngest clusters. In particular, in a number of cases, ages of $\tau >10^7$~yr are assigned to clusters that are in fact HII regions. If the youngest data point in the $M_{\max}$ relation determined by GB08 is removed, or if they had chosen somewhat different (larger) bins to accomodate the systematic errors in the Hunter et~al.\ (2003) age estimates, GB08 would have found a slope of $\delta\approx0.3$--0.4, nearly  indistinguishable from the results presented here. This discrepancy between the original GB08 result and that shown in Figure~4 based on the same dataset highlights the sensitivity of the  $M_{\max}$ method to relatively minor differences in the dating procedure, choice of bins, and the accidental presence or absence of only a few clusters, even assuming that all of the underlying assumptions are true (i.e., constant power-law mass function, no upper mass cutoff, etc.).

\section{IS \boldmath{$M_{\max}$} DUE TO STATISTICS OR TO PHYSICS?}

The $M_{\max}$ method for estimating $\chi(\tau)$, as presented in GB08 and reviewed in Section~4, {\em assumes} that the mass function is a pure
power law with no cutoff at the high-mass end. Several recent works have suggested, however, that there may be a physical (i.e., non-statistical) upper mass limit or cutoff $M_C$ for young star clusters (e.g., McKee \& Williams 1997; Larsen 2009). The goals of this section are two-fold: (1)~to determine whether or not the upper end of the mass function of clusters in the LMC is better fitted by a pure power law or requires a high-mass cutoff;
and (2)~to assess the impact that a physical cutoff would have on the validity of the $M_{\max}$ method for estimating the power law index $\gamma$ for the age distribution.

Perhaps the most familiar example of a population with a physical cutoff is 
the distribution of galaxy luminosities and masses, which is typically described by a Schechter function, $\psi(M) \propto M^{\beta} \mbox{exp}(-M/M_C)$. In this case, the upper end of the mass function has substantial curvature, with the observed number of high-mass galaxies dropping faster
than any power law. To our knowledge, no population of {\em young} clusters
shows definitive curvature at high masses (ancient globular clusters in at least some galaxies, however, do show curvature; Burkert \& Smith 2000; Fall \& Zhang 2001; Jordan et~al.\ 2007). Instead, claims for an upper cutoff
have come primarily from extrapolating the observed power-law distribution beyond the mass of the most massive cluster, to see if clusters are predicted to exist where none is observed.

The mass distributions of young clusters in the LMC ($\tau \lea 10^9$~yr), shown in Figure~7 of CFW10, do not have any obvious down-turn at the
high-mass end. To assess quantitatively whether there is any evidence for an
upper mass cutoff, we integrate the best fit power-law for the mass distribution of clusters with ages $\tau \leq 10^9$~yr and masses $M\geq 3\times10^3~M_{\odot}$, from $M_{\max}$ ($\approx2\times10^5~M_{\odot}$) to infinity. This integration predicts that there should be $\approx3$--4
clusters with masses higher than $2\times10^5~M_{\odot}$. If we use the steepest value of $\beta$ allowed by the fit (rather than the best value), the extrapolation predicts $\approx2\!$ clusters with $M > M_{\max}$. This result may indicate marginal evidence for an upper mass cutoff in the LMC,
although with very low statistical confidence. Larsen (2009) found a similarly ambiguous result for the LMC based on the smaller cluster sample from Bica et~al.\ (1996). We note that if these results do imply a cutoff of $M_C\approx2\times10^5~M_{\odot}$ in the LMC, it must be different from that in the Antennae, where we found that $M_C$ exceeds $10^6~M_{\odot}$ (WCF07; FCW09).\footnote{An upper mass cutoff, regardless of its precise value, may also be needed to account for the absence of very massive, old clusters in the LMC (i.e., those with $M\gea 10^6~M_{\odot}$ and $\tau \gea 10^9$~yr).}

A physical upper cutoff in the mass function would invalidate, or at least complicate, estimates of the age distribution based on the $M_{\max}$ method descibed in Section~4. While it is straightforward to replace a power-law mass function, $\psi(M) \propto M^{\beta}$, by a Schechter
function, $\psi(M) \propto M^{\beta}~ \mbox{exp}(-M/M_C)$ and to revise Equations~(3), (4), and (5) for $M_{\max}$ accordingly, the formula for $\delta$ would then involve three parameters: $\beta$, $M_C$, and $\gamma$.
Thus, to estimate the exponent of the age distribution $\gamma$ from $\delta$, one would need to know both $\beta$ and $M_C$ and whether or not they depend on age. But this information would in turn essentially require a determination of the bivariate distribution $g(M, \tau)$ and thus a direct determination of the age distribution in the first place (the method we advocate). In practice, therefore, the $M_{\max}$ method and the claim of an upper cutoff in the mass function of clusters are incompatible.

\section{IS THE AGE DISTRIBUTION OF STAR CLUSTERS ``UNIVERSAL''?}

We have shown here that different methods, both direct and indirect, give the same declining power-law shape, $\chi(\tau) \propto \tau^{\gamma}$
with $\gamma\approx-0.8$, for the age distribution of star clusters in the LMC. We have previously interpreted this declining shape as due primarily to the disruption rather than to the formation of the clusters, and also suggested that the shape of the age distribution reflects a combination of several different disruption processes, rather than any single process. The two disruption mechanisms that are most likely to dominate on short timescales ($\tau \lea 10^8$~yr), expulsion of interstellar material by stellar feedback and mass loss due to stellar evolution, are processes internal to the clusters themselves and are not sensitive to the external environment (see FCW09 and Fall et~al.\ 2010 for a more detailed discussion of early disruption
processes). This implies that different galaxies may have cluster age distributions that are broadly similar to that observed in the LMC, at least for the first $\sim\!10^8$~yr and possibly the first $\sim\!10^9$~yr.

There is now growing evidence to support this hypothesis. In the following, we summarize recent results for the age distributions of young star clusters in more than a dozen nearby galaxies. In virtually all of these galaxies, the age
distribution has a declining form similar to that shown here for the LMC, although the quality of the data and level of analysis vary.

In a series of papers, we have made a detailed study of massive ($M\gea 10^4~M_{\odot}$) clusters in the merging Antennae galaxies based on deep \textit{UBVI} H$\alpha$ images taken with the WFPC2 camera on \textit{HST}, and more recently from deeper, higher-resolution images taken with the \textit{HST}/ACS camera. The masses and ages of the clusters, as well as the completeness of the sample, have been fully quantified (e.g., Whitmore et~al.\ 1999; FCW05; WCF07; FCW09; Whitmore et~al.\ 2010). From the direct counting method, we found that the (mass-limited) age distribution  can be approximated by a power-law with $\gamma\approx-1$ for $\tau \lea 10^9$~yr (FCW05; WCF07; FCW09), and has a similar shape, including the peak at $\tau\lea10^7$~yr, for large regions within the Antennae that are separated by distances of nearly $\approx$10~kpc. There are no hydrodynamical processes that could synchronize a burst of cluster formation this precisely over such large separations (WCF07; FCW09; Whitmore et~al.\ 2010), indicating that the disruption rather than the formation of the clusters is primarily responsible for the observed shape of the age distribution.

In CFW10, we constructed the age distribution of clusters in the SMC using the direct counting method, based on the sample and \textit{UBVR} photometry provided by Hunter et~al.\ (2003). We found that the age distribution of clusters in the SMC is similar to that for the LMC, with $\gamma\approx-0.8$ for $\tau \lea 10^9$~yr, although the statistics are poorer and the results less certain than for the LMC, particularly for $\tau \lea 10^7$~yr, where current samples of clusters are incomplete (see the Appendix in CFW10 for more details).

Lada \& Lada (2003) compiled a catalog of star clusters in the solar neighborhood. Although the distributions are noisier than for the Magellanic Clouds, their Figure~3 shows an age distribution of the form $dN/d\log\tau\approx\mathrm{const}$ for embedded and non-embedded clusters with ages $10^6 \lea \tau \lea 10^8$~yr and masses $M\lea10^3~M_{\odot}$. This is equivalent to $\chi(\tau) \propto \tau^{-1}$, similar to our results for the LMC.

Mora et~al.\ determined the age distribution of star clusters in four nearby spiral galaxies, NGC~1313, NGC~4395, NGC~5236 (M83), and  NGC~7793, based on broad-band images taken with the ACS and WFPC2 cameras on \textit{HST}. They found that the age distributions all decline steeply 
by counting clusters brighter than a given $V$-band luminosity, consistent with $\approx\!80$\% of the clusters in a given mass interval being disrupted every decade in age for $\tau \lea 10^9$~yr (after converting their luminosity-limited results to mass-limited ones). In the notation used here,
this is equivalent to $\gamma \approx-0.7$. Our recent analysis of 
multi-band images of a field in M83 taken with the newly installed WFC3 camera on \textit{HST} supports the Mora et~al.\ result for this galaxy
(Chandar et~al., in prep.). The $M$--$\tau$ diagram shows approximately equal numbers of clusters in equal bins of $\log\tau$ above a given mass,
and the (mass-limited) age distribution from the direct counting method gives $\gamma\approx-0.9\pm0.2$ for clusters with $\tau \lea \mbox{few}\times 10^8$~yr and $M\gea \mbox{few} \times 10^3~M_{\odot}$. In contrast, GB08 {\em inferred} $\gamma\approx0$ from the $M_{\max}$ method for M83, based on an estimated slope $\delta\approx1$ and the
assumption $\beta=-2.0$. The GB08 result, however, is dominated by a single data point, the youngest one, where they find a maximum mass of only $M \sim 10^3~M_{\odot}$ for $\tau < 10^7$~yr clusters, reminiscent of their result for the LMC. The \textit{HST}/WFC3 data clearly show several 
clusters with $M\sim10^5~M_{\odot}$ and $\tau \lea 10^7$~yr in M83
(i.e., they are very luminous HII regions), indicating that the data used by GB08 either have systematic errors in the age and mass estimates or are flawed in some other way.

Pellerin et~al.\ (2010) estimated the ages and masses of star clusters in the nearest collision ring galaxy, NGC~922, from broad-band \textit{HST}/WFPC2 images. Their Figure~8 shows that the cumulative age distribution of clusters more massive than $10^5~M_{\odot}$ is reasonably well matched by a model with $\gamma\approx-0.6$ for $\tau \lea 10^8$~yr. The formation rate of clusters, meanwhile, is predicted to have decreased
over the same period of time, based on $N$-body/SPH simulations of the
collision. This indicates that it is the disruption rather than the formation
of the clusters that is primarily responsible for the observed shape of the age distribution in NGC~922, similar to our conclusions for the LMC and the Antennae.

Goddard et~al.\ (2010) determined the age distribution of clusters in 
NGC~3256, a pair of gas-rich galaxies that are further along in the merging process than the Antennae, based on \textit{UBVI} images taken with the
\textit{HST}. Their Figure~5 shows that the age distribution declines like a power law for clusters more massive than $M \approx 5\times10^5~M_{\odot}$ and younger than $\tau \lea 2\times10^8$~yr. Similarly, the $M$--$\tau$ diagram shown in their Figure~4 has approximately equal numbers of clusters
in equal bins of $\log\tau$ in different intervals of mass.

Peterson et~al.\ (2009) estimated ages and masses of star clusters in the interacting galaxy pair Arp~284, from broad-band \textit{HST}/WFPC2 images. Although there are relatively few clusters, the $M$--$\tau$ diagram shown in their Figure~15 has approximately equal numbers of clusters
in equal bins of $\log\tau$ for $M \gea \mbox{few}\times10^5~M_{\odot}$,
indicating that the age distribution declines with $\gamma\approx-1$.

We have recently determined ages and masses for compact star clusters in the spiral galaxy M51, from multi-band (\textit{UBVI} H$\alpha$) images
taken with the ACS and WFPC2 cameras on \textit{HST}. The $M$--$\tau$ diagram from our dating analysis is similar to that shown here in Figure~3 for the LMC, with the number of clusters increasing slowly in equal bins of $\log\tau$ for ages $\tau \lea \mbox{few}\times10^8$~yr and masses $M\gea10^4~M_{\odot}$ (Chandar et~al., in prep). The $M$--$\tau$ diagram indicates that the age distribution of clusters in M51 declines in a fashion
similar to that for clusters in the LMC.

Melena et~al.\ (2009) estimated the ages and masses of star clusters and complexes in nine nearby, star-forming dwarf galaxies, selected from \textit{GALEX} near-ultraviolet images and measured at ultraviolet, optical, and 
near-infrared wavelengths. These regions have an age distribution that 
declines as $\chi(\tau) \propto \tau^{-1}$ for mass-limited samples, similar to
the results described above, although the larger apertures used in the
Melena et~al.\ study may include more than one compact cluster or association.

Taken together, these results are quite striking. They suggest that the age distribution of star clusters in over a dozen galaxies, which include dwarf and giant galaxies, isolated and interacting galaxies, irregular and spiral galaxies, all show a similar, declining shape. Furthermore, the age distributions have been constructed for clusters covering nearly four orders of magnitude in mass, from $\sim\!10^2~M_{\odot}$ to $\sim\!10^6~M_{\odot}$. These results strongly support our hypothesis that the age distributions of young cluster systems are approximately ``universal.''

Our explanation for this is that the age distribution is dominated by the disruption rather than the formation of the clusters. It is far more likely that the clusters in all of these galaxies have similar disruption histories than it is that they have similar formation histories and that we also happen to be observing them all at the same special time when their formation rates just happen to have peaked within the past $10^7$~yr, i.e., within 0.1\% of the Hubble age. Of course, variations in the cluster formation rate {\em will}
affect the observed age distributions of the clusters, even if the disruption rate is identical in different galaxies. In other words, we do expect some variations in $\gamma$ among different galaxies, with higher or lower values
depending on whether the rate of cluster formation increased or decreased over the past $\sim\!10^8$ or $10^9$ years.\footnote{Even some interacting galaxies will have below-average $\gamma$ if the peak rate of cluster formation occurred at any time other than the present.} The observational evidence so far, however, suggests that variations in $\gamma$ are relatively modest. For a sample of randomly selected galaxies, we expect that the average rate of cluster formation will be approximately constant; thus, the average value of $\gamma$ for the whole sample should be the true disruption exponent.

Our picture of (more or less) continuous disruption of star clusters starting soon after birth, and operating roughly independent of mass, has important implications for the origin and location of stars within galaxies. Early disruption naturally reconciles the fact that most stars form in clusters (some bound, some unbound) as defined here (e.g., Carpenter 2000; Lada \& Lada 2003; FCW05), yet most stars reside in the field. In other words, stars from dispersed clusters constitute most of the field stellar population.
Without disruption, i.e., $\chi(\tau) \approx \mbox{const}$ or $\gamma \approx0$, most stars would have to form in the field regions of galaxies, rather than in clusters, which is not observed. Early disruption of clusters also affects the observed shape of the age distribution of field stars, which can appear decoupled from that of the clusters after only $\sim\!10^7$~yr, since $\sim\!80$\% of clusters have already dispersed their stars to the
field by this time (Chandar et~al.\ 2006).

\section{SUMMARY AND CONCLUSIONS}

In this paper, we used two different methods to determine the exponent $\gamma$ of the age distribution of star clusters in the LMC. The first method, 
which we have used previously in our studies of clusters in the Antennae galaxies and in the Magellanic Clouds, is based on counting clusters directly in the mass-age plane. In effect, we first determine the joint distribution of masses and ages $g(M, \tau)$ and then integrate over bands in mass,
although we use fairly large bins in $\log\tau$ to accomodate the systematic age uncertainties which arise when comparing integrated colors with stellar population models. The second method for estimating $\gamma$, advocated by Gieles \& Bastian (2008), is based on the maximum masses of clusters as a function of their ages, and is referred to as the $M_{\max}$ method. This method uses only a small fraction of the data (the upper envelope of the $M$--$\tau$ distribution), and also requires extra assumptions about the form of the mass function. We found that, for the clusters in the LMC, the (indirect)
$M_{\max}$ method gives the same result as that from our direct counting method, namely $\gamma\approx-0.8$  for $\tau \lea 10^9$ yr. In contrast, Gieles \& Bastian inferred $\gamma\approx0$ from the $M_{\max}$ method, because their youngest data point was artificially low. This highlights the sensitivity of the $M_{\max}$ method to relatively minor differences the age dating procedure, such as the adopted stellar population models, details of the extinction correction, and/or the accidental presence or absence of only a few clusters (e.g., R136).  We strongly advocate the direct counting method, which has the fewest assumptions and 
makes use of all the clusters in a sample.

We compiled results from the literature for the age distributions of young star clusters in more than a dozen nearby galaxies, including dwarf and giant,
isolated and interacting, irregular and spiral galaxies. The distributions all have declining shapes, similar to that found here and in CFW10 for clusters in the LMC. These results support our hypothesis that the age distributions of
young cluster systems in nearby galaxies have an approximately ``universal'' shape which mainly reflects the disruption rather than the formation histories of the clusters.

\acknowledgements
We thank Dean McLaughlin, Francois Schweizer, Soeren Larsen, and the anonymous referee for helpful comments. SMF acknowledges support from the Ambrose Monell Foundation and from NASA grant AR-09539.1-A. 
RC acknowledges support from NASA grant GO-10402.11-A, awarded by the Space Telescope Science Institute, which is operated by AURA, Inc., under NASA contract NAS5-26555.

\begin{figure}
\epsscale{0.7}
\plotone{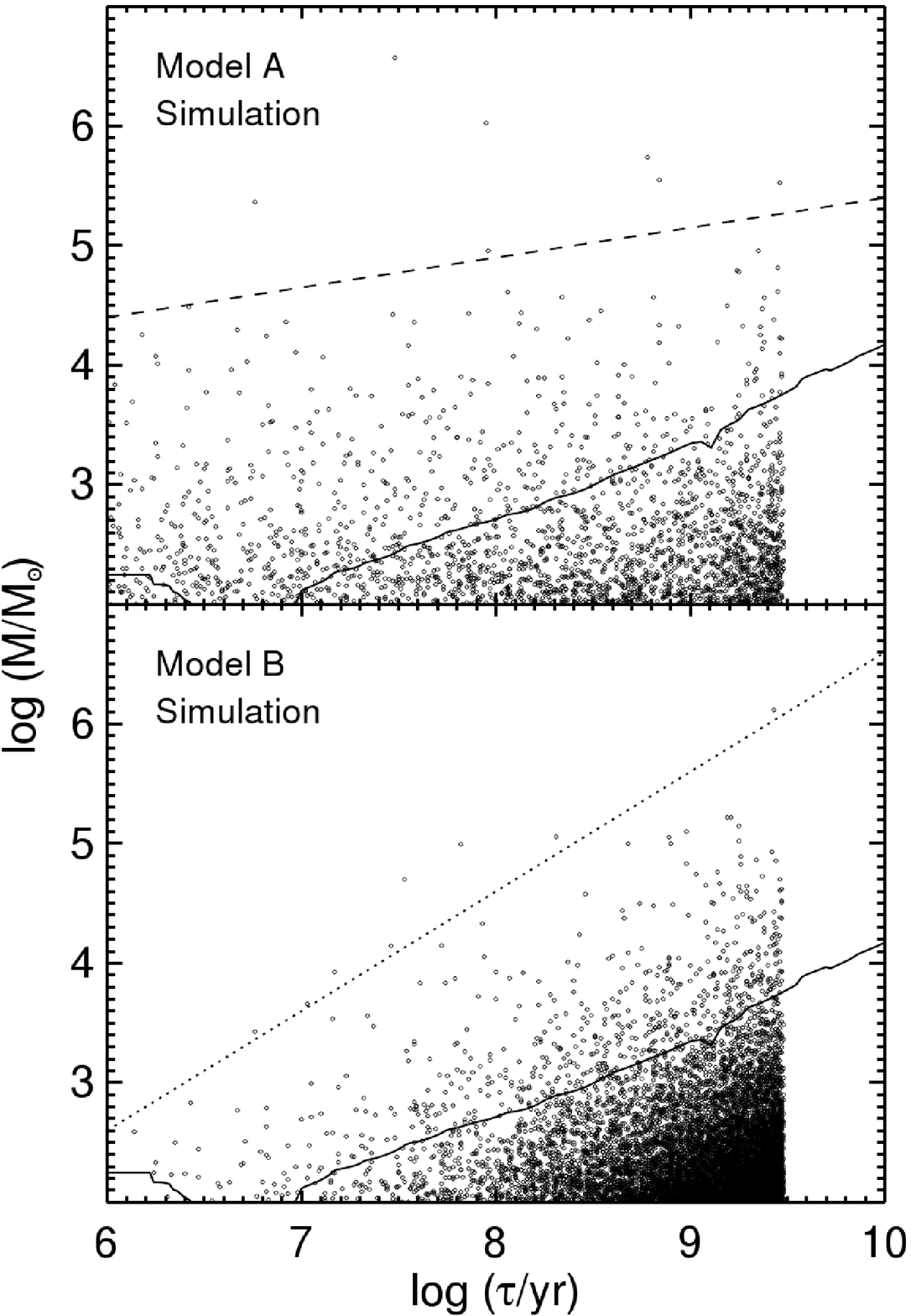}
\caption{
Simulated $M$--$\tau$ diagrams from Monte Carlo realizations of Model~A with $g(M,\tau) \propto M^{-1.8}\tau^{-0.8}$ (top panel) and Model~B with $g(M,\tau) \propto M^{-2.0}\tau^{0.0}$ (bottom panel). The thick solid line in each panel represents $M_V=-4.0$. The simulations of both models have equal numbers of clusters above the lines. The dashed line shows the predicted $M_{\max}$ relationship for Model~A, and the dotted line shows the predicted relationship for Model~B. See text for details.
}
\label{fig:monte}
\end{figure}

\begin{figure}
\epsscale{0.7}
\plotone{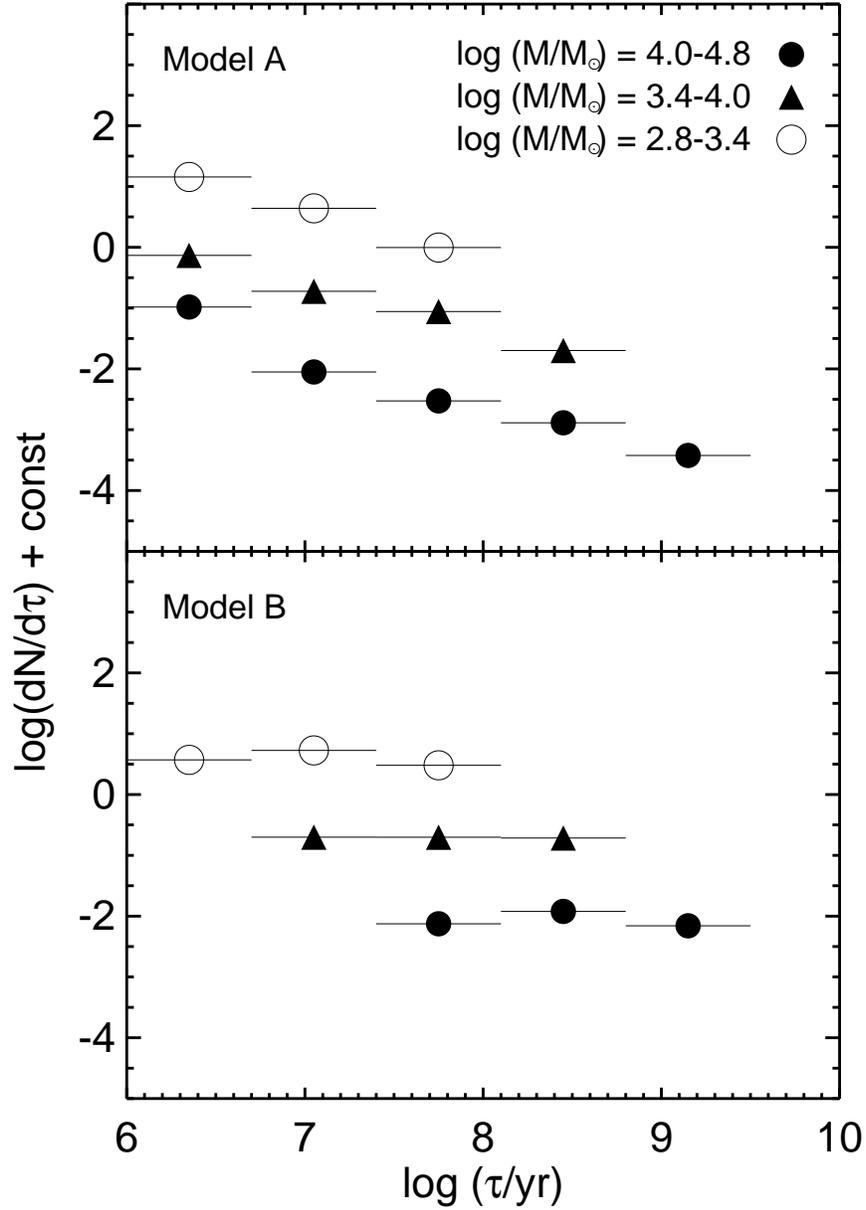}
\caption{Predicted age distributions $\chi(\tau)=dN/d\tau$ from Monte Carlo realizations of Model~A (top panel) and Model~B (bottom panel) for the indicated intervals of mass. Three data points are missing at young ages for Model~B because no clusters were predicted with these masses and ages.
}
\label{fig:dndt}
\end{figure}

\begin{figure}
\epsscale{0.7}
\plotone{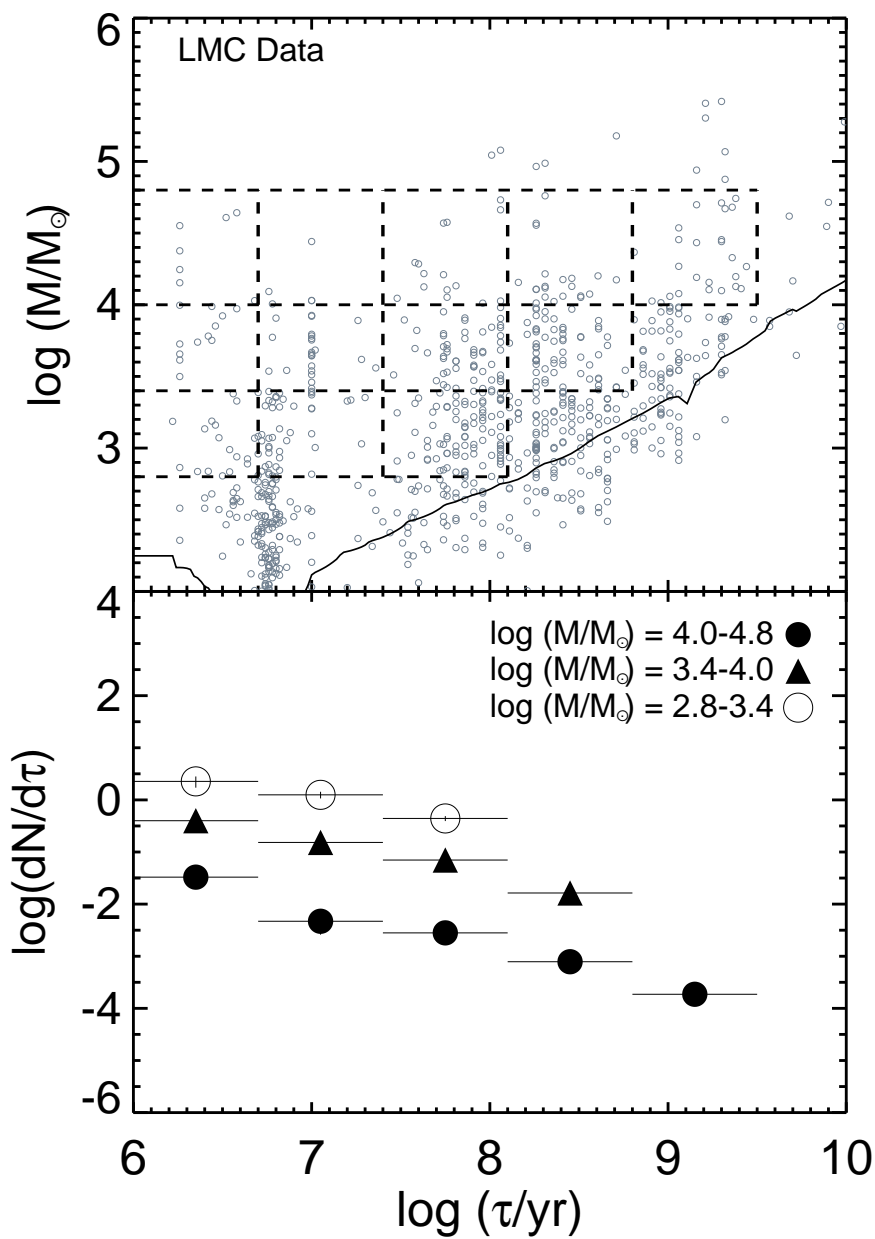}
\caption{Observed $M$--$\tau$ diagram of clusters in the LMC (top panel), based on mass and age estimates from CFW10. The dashed rectangles show the $M$--$\tau$ bins used to construct the cluster age distributions shown in the bottom panel for the indicated intervals of mass (the same as in CFW10).
}
\label{fig:Mt}
\end{figure}

\begin{figure}
\epsscale{1.0}
\plotone{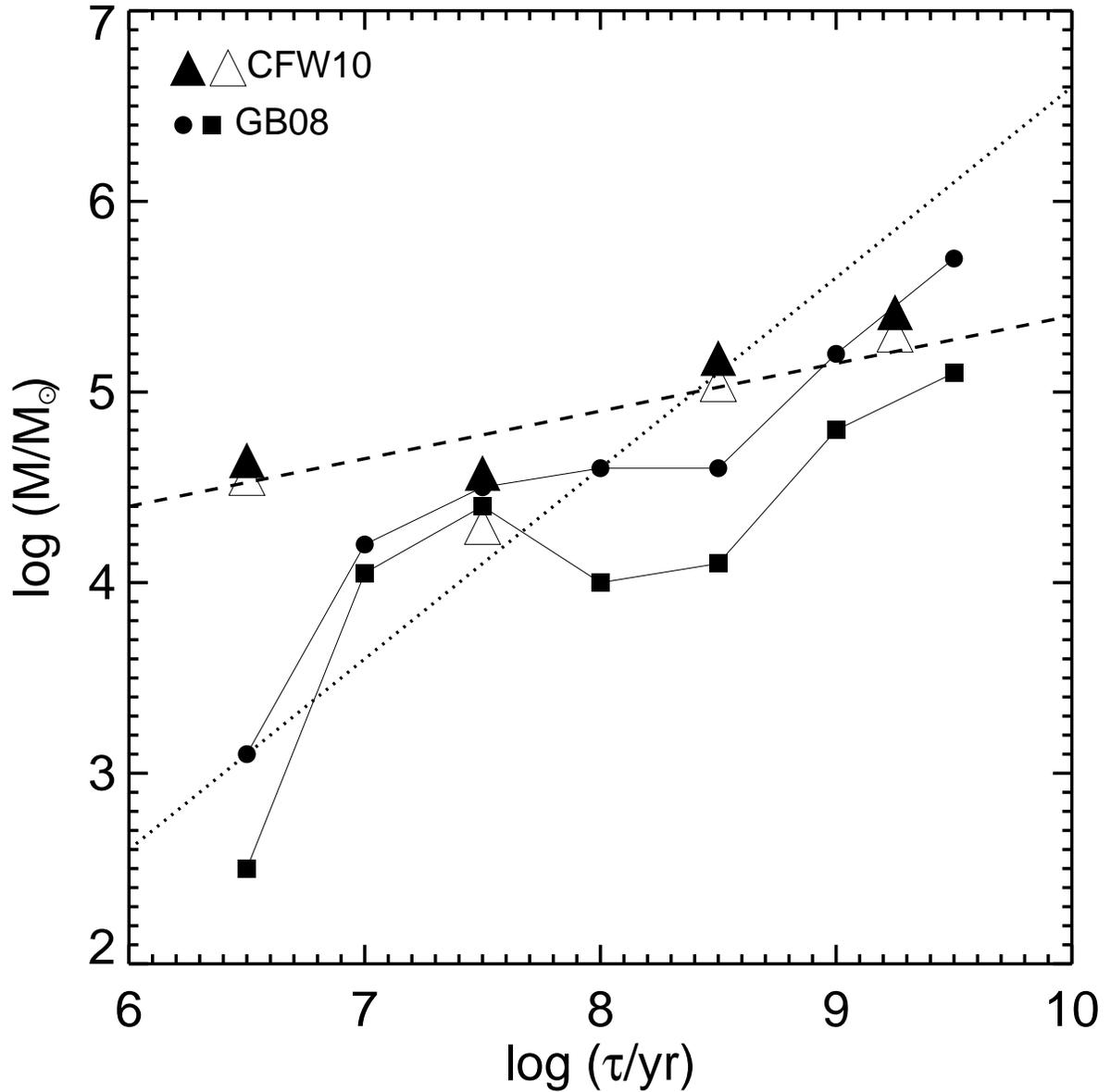}
\caption{The $M_{\max}$ relation for the first and third most massive clusters in the LMC based on our age and mass estimates for clusters in the Hunter et~al.\ sample (filled and open triangles, respectively). The small circles and squares show the relationship found by Gieles \& Bastian (2008) for the first and third most massive clusters, respectively. The dashed and dotted lines are the same as shown in Figure~1, and represent predictions
from Models~A and B. See text for details.
}
\label{fig:Mmaxus}
\end{figure}

\begin{figure}
\epsscale{1.0}
\plotone{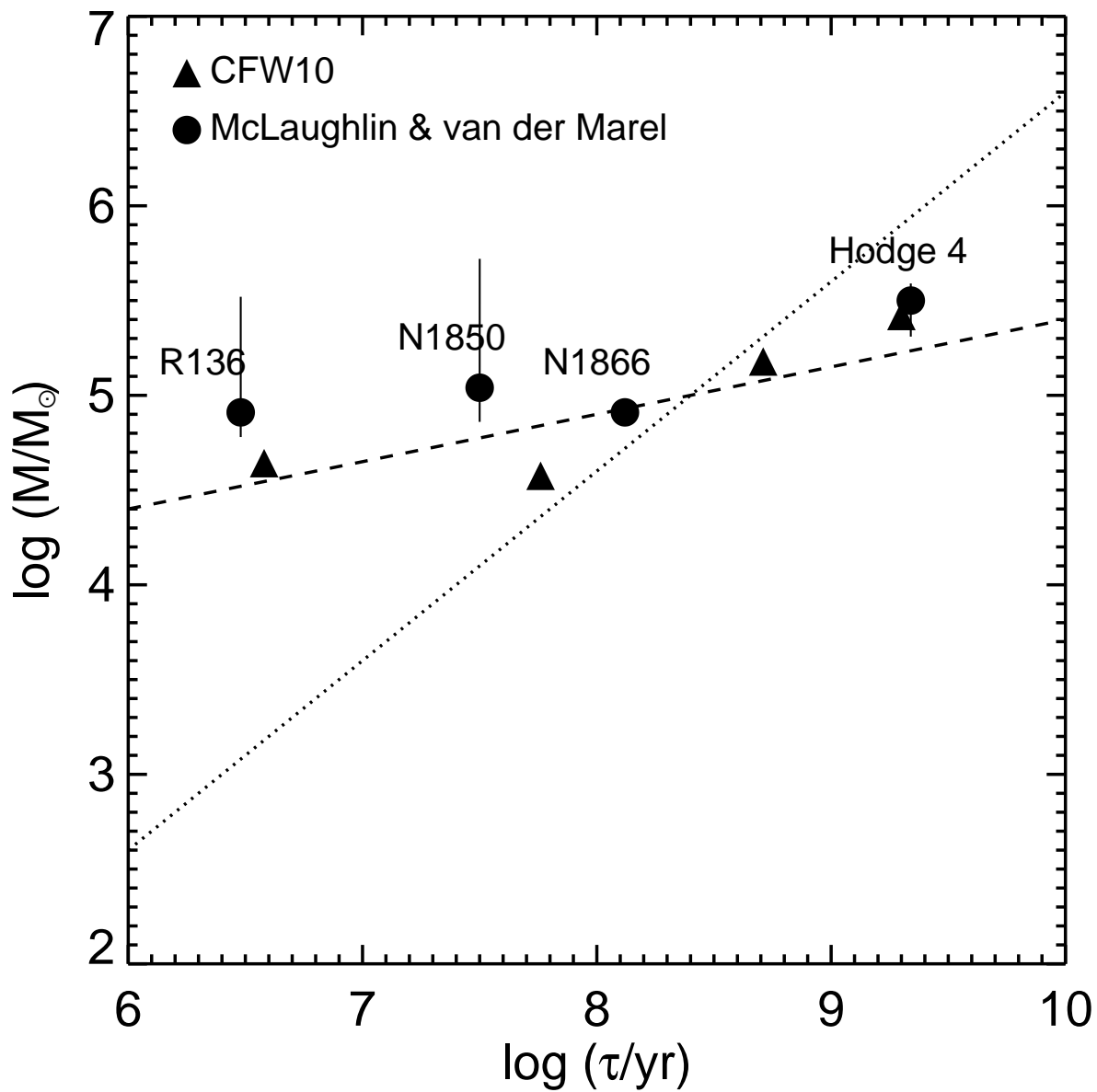}
\caption{The $M_{\max}$ relation for clusters in the LMC based on the sample of McLaughlin \& van der Marel (2005) (filled circles) and the sample of Hunter et~al.\ (2003) with our mass and age estimates (filled triangles).
The names of the clusters from the McLaughlin \& van der Marel sample
are labeled. The dashed and dotted lines are the same as in Figures~1 and 4.
The dashed line, i.e., $\gamma=-0.8$, is clearly preferred.
}
\label{fig:Mmaxdean}
\end{figure}

\end{document}